\documentclass[preprint]{ptephy_v1}

\preprintnumber{XXXX-XXXX} 

\usepackage{amsmath} 
\usepackage{hyperref} 
\usepackage{url} 
\usepackage{natbib}

\newcommand{\nuc}[2]{$^{#1}${#2}}

\begin{document}

\title{The southern shore of the island of inversion studied via quasi-free scattering}


\author{Julian Kahlbow}
\affil{Massachusetts Institute of Technology, Cambridge, MA 02139, USA \email{jkahlbow@mit.edu}}

\begin{abstract}%
Neutron-rich nuclei exhibit a variety of intriguing features associated with nuclear structure evolution, deformation, and other phenomena.
Particularly interesting is the region in the chart of nuclides around $Z=12$ and $N=20$, commonly referred to as ``Island of Inversion'', which is profoundly influenced by these features.
Recent cutting-edge experiments performed at \mbox{SAMURAI}/\mbox{RIBF} have investigated the structure of the most neutron-rich O and F isotopes, including \nuc{27,28}{O} and \nuc{28-30}{F}, utilizing quasi-free scattering and invariant-mass spectroscopy techniques.
This experimental campaign manifests the breakdown of the $N=20$ magicity for O and F isotopes, placing them within the ``Island of Inversion'', as is discussed in this review article.
The results are further supported by theoretical analyses employing state-of-the-art shell-model and ab-initio calculations.
These nuclei serve as corner stones for the study of weak binding and continuum coupling, deformation, and halo formation.
Signatures for the establishment of a superfluid regime in \nuc{28}{O} and \nuc{29}{F} are found.
Future experimental and theoretical studies are needed to examine details.
\end{abstract}

\subjectindex{xxxx, xxx}

\maketitle

\section{Introduction -- The nuclear island of inversion}
Atomic nuclei away from the valley of $\beta$-stability, neutron-rich or -deficient, present a variety of new and surprising phenomena that nuclear physics tries to understand and describe based on the underlying nuclear interaction.
For extremely neutron-rich systems with up to  $\sim 300$\% more neutrons than protons, new questions come up as to the state and structure of such very isospin-asymmetric quantum many-body systems.
A region in the chart of nuclides that attracts a lot of attention in that regard is called ``Island of Inversion'' (IoI) around proton number $Z=12$~\cite{warburton90,caurier14,otsuka20}.
Nuclei considered part of the IoI are characterized by a vanishing neutron $N=20$ magic number, large occupation of two-particle-two-hole ($2p2h$) excitations, and deformed ground states.
In a shell-model picture, the interplay between correlation energy gains and weakening monopole interaction leads valence neutrons to energetically favor to occupy the $pf$ shell instead of $d_{3/2}$ orbital.

A prime example for characteristics of the IoI are the neutron-rich Mg ($Z=12$) isotopes.
The $N=20$ \nuc{32}{Mg} was found to have a deformed ground state~\cite{wimmer10}, similar behavior becomes apparent for Mg isotopes approaching $N=28$, in which the $N=28$ is predicted to vanish for \nuc{40}{Mg} in favor of a deformed ground state~\cite{caurier04}.
Accessing the \nuc{40}{Mg} nucleus is experimentally the front line and has recently been measured and studied for the first time via in-beam $\gamma$-ray spectroscopy at RIBF~\cite{crawford19,crawford14}, unveiling unexpected structure. 
Meanwhile, more such regions with similar behavior and strong nuclear structure and shell evolution have been identified~\cite{nowacki21} throughout the neutron-rich side of the chart of nuclides around supposedly magic numbers, often using $\gamma$-ray spectroscopy as first experimental evidence as discussed in other review articles of this topical collection. 

The IoI has been established at an upper proton number $Z=13$~\cite{heylen16} down to $Z=10$, i.\,e., when the $\pi(0d_{5/2})$ orbital is not fully occupied.
However, the lower bound has long been unknown as it is experimentally very challenging to study the most neutron-rich fluorine ($Z=9$) and oxygen ($Z=8$) isotopes approaching the drip line as many are neutron-unbound or difficult to access because of low production rates.
On the other hand, those nuclei are very interesting to study as they are entering a weakly-bound regime adding continuum degrees of freedom and pushing limits of nuclear theory for both classical shell-model and ab-initio approaches.
Theories predict a number of effects in this region of the chart of nuclides such as shell evolution, two-neutron halo nuclei, and an enhanced degree of collectivity, thus being particularly sensitive to nuclear theory and benchmarking nuclear interactions and many-body methods.  

This review focuses on latest experimental results from \mbox{RIBF}/\mbox{RIKEN} (Japan) combined with theoretical results aiming at the study of \nuc{28}{F}--\nuc{30}{F} and \nuc{27,28}{O} in which quasi-free scattering is used to populate these unbound nuclear systems and perform their single-particle spectroscopy.
This article discusses in particular the experimental works in Refs.~\cite{kondo23,revel20,kahlbow24}, additional results from the same experimental campaign on the neon isotopes can for example be found in Refs.~\cite{holl22,wang23}.

\section{Experimental procedure}
\label{sec:exp}
\begin{figure}[]
\centering\includegraphics[width=0.95\textwidth]{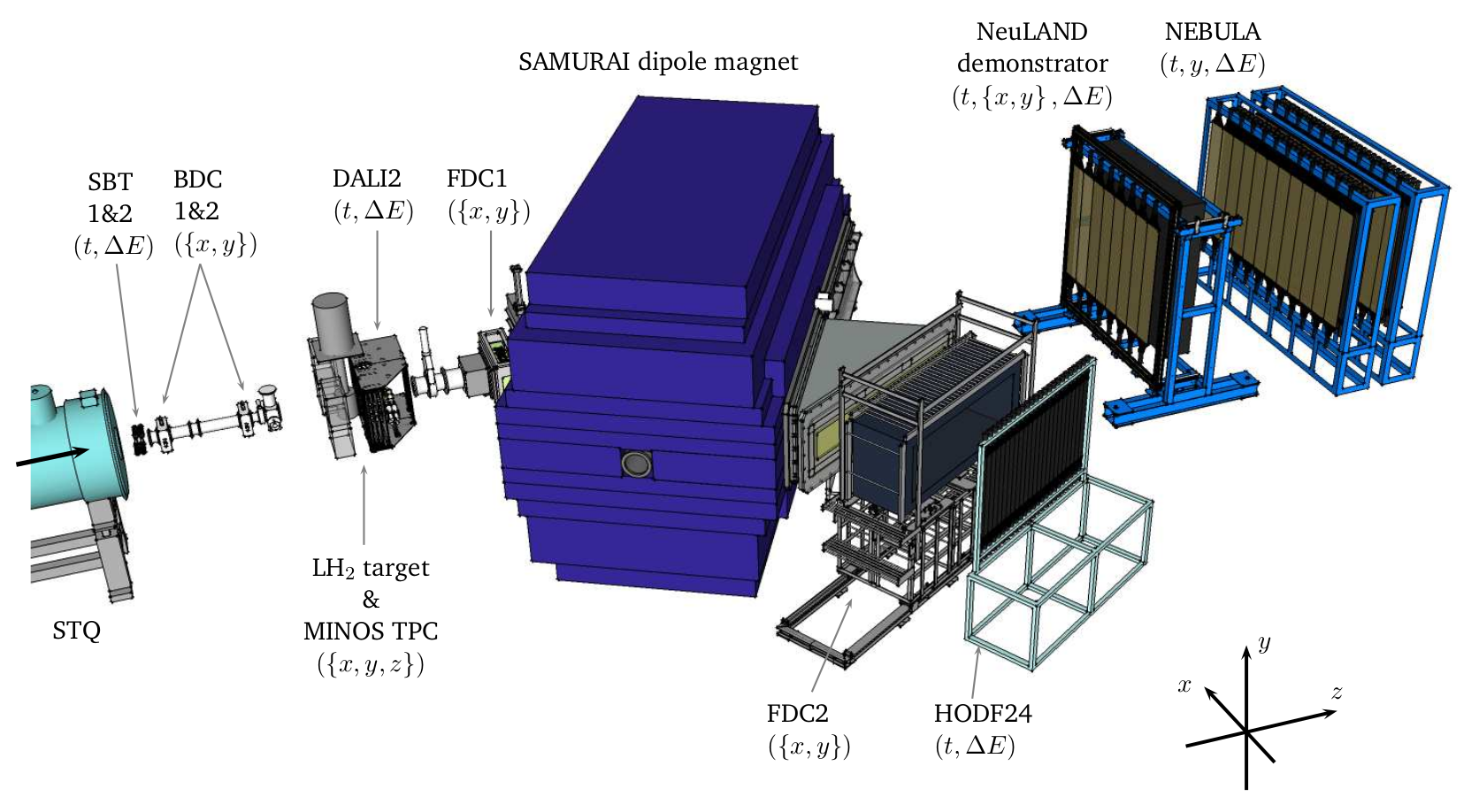}
\caption{\mbox{SAMURAI} experimental setup for the invariant-mass spectroscopy with RI beams. The unique combination of the \mbox{MINOS} LH$_2$ target and TPC together with the neutron detectors \mbox{NeuLAND} demonstrator and \mbox{NEBULA} leads to unprecedented luminosity and resolution. }
\label{fig:samurai_setup}
\end{figure}

The experimental results discussed in this review were obtained in an experimental campaign conducted at the \mbox{SAMURAI} setup at RIKEN's Radioactive Isotope Beam Factory (\mbox{RIBF})~\cite{kobayashi13}.
Quasi-free nucleon knockout reactions $(p,pN)$ are employed to either populate neutron-unbound ground-states, bound- and unbound-excited states, or perform the single-particle spectroscopy, leveraging the sudden nucleon removal and thus accessing the momentum measurement of the spectator $A-1$ system via $\vec{p}_{A-1}=-\vec{p}_N$~\cite{jacob66,hansen03,aumann13}.

This topical collection discusses the unique opportunities provided by utilizing quasi-free scattering (QFS) reactions on the \mbox{MINOS} LH$_2$ target device coupled with a time-projection chamber (TPC)~\cite{obertelli14}, which significantly enhanced experimental performance and data quality at the \mbox{SAMURAI} setup.
The \mbox{SAMURAI} setup enables kinematically complete measurements in inverse kinematics, detecting all reaction particles, including charged particles, $\gamma$-rays, and fast neutrons.
The integration of \mbox{MINOS} with the neutron detectors \mbox{NeuLAND} demonstrator~\cite{boretzky15,boretzky21} and \mbox{NEBULA}~\cite{kobayashi13,nakamura16} resulted in unprecedented performance in invariant-mass spectroscopy with radioactive-ion (RI) beams, improving resolution by a factor of about four and increasing luminosity by about a factor of $1,000$ compared to similar experiments in the past~\cite{christian12}.
The experimental setup used in the presented campaign is shown in Fig.~\ref{fig:samurai_setup}.
The $20$\,cm thick LH$_2$ target increases the reaction probability, while the vertex tracking using the surrounding TPC in $(p,pN)$ reactions restores resolution.
Together with high-resolution and high-efficiency neutron detector arrays, this setup enables the study of previously unknown nuclear structure, as demonstrated for \nuc{28,29}{F}, and facilitates discovery measurements as seen in the cases of the neutron-unbound \nuc{30}{F} and \nuc{27,28}{O}.

In addition to in-flight $\gamma$-ray spectroscopy utilizing the \mbox{DALI2} NaI(Tl) array~\cite{takeuchi14}, the presented experiments take advantage of neutron spectroscopy using granular plastic-scintillator arrays.
Adding the \mbox{NeuLAND} demonstrator~\cite{boretzky21}, a detector at the R$^3$B setup at \mbox{GSI-FAIR}, to the existing \mbox{NEBULA} array~\cite{nakamura16} nearly doubles the effective interaction depth.
The single-neutron detection efficiency of the \mbox{NeuLAND} demonstrator was assessed at \mbox{SAMURAI} using quasi-monoenergetic neutrons at $250$\,MeV, produced in a charge-exchange reaction $^7$Li$(p,n)^7$Be(g.s.+$0.43$\,MeV), incident on the $40$\,cm thick detector.
The obtained efficiency of $27$\% (for $\Delta E>5$\,MeVee) agrees within $1$\% with simulations~\cite{kahlbow24} and is similar to results obtained for \mbox{NEBULA}~\cite{kondo20}.
This translates to a reconstruction efficiency for single-neutron decays of approximately $54$\% for a relative energy up to $1$\,MeV, decreasing to about $30$\% at a relative energy of $3.5$\,MeV due to acceptance effects.
The measurement of multi neutrons is exacerbated by the creation of secondary-particle showers in the detector material, necessitating the application of so-called crosstalk-rejection algorithms to suppress the reconstruction of false neutron hits~\cite{nakamura16,kondo20}. 
This further reduces the reconstruction efficiency, resulting in a two-neutron efficiency of no more than $15$\%~\cite{kahlbow19}.
The efficiency decreases further for three- and four-neutron signals to no more than $2$\% and $0.4$\%, respectively~\cite{kondo23}.   
The relative energy resolution is as good as $\sigma\sim 100$\,keV at $E_{rel}=1$\,MeV and better than $30$\,keV at $0.1$\,MeV, reaching values comparable to in-beam $\gamma$-ray spectroscopy with scintillator arrays.

In addition to the detector setup, the high RI beam intensities at the \mbox{RIBF} enable experiments at the drip lines with increased count rates.
For instance, the measurement of \nuc{28}{O} took advantage of the high-intensity \nuc{29}{F} secondary beam with approximately $90$\,counts per second produced through fragmentation and in-flight separation at the \mbox{BigRIPS} separator~\cite{kubo07}, using a \nuc{48}{Ca} primary beam at an unprecedented current of around an average of $600$\,pnA at that time.
This underscores the power and uniqueness of combining state-of-the-art experimental techniques leading to unprecedented results from \mbox{QFS} experiments, which are discussed in the subsequent sections.

\section{The neutron-rich F isotopes}
\label{sec:f_isotopes}
Only recently the neutron drip-line for the F isotopes has been experimentally determined, revealing that \nuc{31}{F} is the last bound F isotope~\cite{ahn19,ahn22,sakurai99}.
Notably, the isotopes \nuc{28,30}{F} are unbound by neutron decay.
The \nuc{29}{F} nucleus consists of $20$ neutrons and one proton above the purpotedly closed $Z=8$ magic proton shell.

There are indications suggesting that the $N=20$ neutron magic number is broken in \nuc{29}{F} supported by the observation of a low-lying bound excited state at $1.08$\,MeV through $\gamma$-ray emission~\cite{doornenbal17}.
This observation aligns with earlier predictions from \mbox{SDPF}-space shell-model calculations for the \mbox{IoI}, indicating almost degenerate $sd$ and $pf$ shells.
Moreover, calculations included in~\cite{doornenbal17}, that purposefully do not account for a correlation-energy gain, i.\,e., a large $N=20$ gap in a restricted \mbox{USDB} model space, predict a single-particle excitation of at least $3$\,MeV for the first excited state, contradicting the experimental finding.
The predicted spin-parity for \nuc{29}{F} is $5/2^+$ with the first excited state assumed to be $1/2^+$, stemming from a multiplet coupling a \nuc{28}{O}($2^+$) state to the valence proton $\pi d_{5/2}$.
Also, the mass measurement of \nuc{29}{F} and its extrapolated trend of $S_{2n}$ values suggest a departure from shell closure~\cite{gaudefroy12}, as well as the matter-radius measurement indicating a $2n$ halo~\cite{bagchi20}. 

The onset of shell evolution in the F isotopic chain remains ambiguous, particularly as an earlier measurement of the unbound \nuc{28}{F} suggested that identified states~\cite{christian12}, including the ground state and possibly an excited state, could indeed be described by shell-model calculations using the restricted \mbox{USDB} interaction, implying a more traditional shell structure.
Such a conclusion would also raise doubts about the structural description of \nuc{29}{F}.
However, this previous study suffered from low statistics and poor invariant-mass resolution~\cite{christian12}, while the $\gamma$-ray spectroscopy of \nuc{27}{F} already shows an indication for shell-structure evolution~\cite{elekes04,doornenbal17}.
More nuclear-structure results from proton-knockout reactions along the F isotopic chain are for example discussed in Ref.~\cite{vandebrouck17}. 

\subsection{\nuc{28}{F} and \nuc{29}{F}}
The existing limited results of \nuc{28,29}{F} demand a more detailed and conclusive study and spectroscopy. 
Analyzing the neutron knockout on \nuc{29}{F} will provide crucial insights into both the structure of the populated \nuc{28}{F} and the valence-neutron structure of the initial \nuc{29}{F}.

In the experimental campaign detailed in Sec.~\ref{sec:exp} the QFS neutron knockout on \nuc{29}{F}(g.s.) was studied populating the unbound \nuc{28}{F}, observed as \nuc{27}{F} alongside a coincident neutron, denoted as \nuc{29}{F}$(p,pn)$\nuc{28}{F}$\rightarrow$\nuc{27}{F}$+n$~\cite{revel20}.
This measurement capitalized on high statistics and unprecedented invariant-mass resolution.
The resulting relative-energy spectrum of \nuc{28}{F} uncovered a complex structure of unbound states, featuring a low-lying ground state at $198(6)$\,keV with width $\Gamma=180\pm40$\,keV, along with certain states decaying through excited states in \nuc{27}{F}~\cite{revel20}.
Furthermore, the QFS proton-knockout on the \nuc{29}{Ne} beam, part of the same experimental campaign, provided a complementary method to populate \nuc{28}{F}, particularly sensitive to negative-parity states.
Both neutron and proton knockouts populated the same ground-state of \nuc{28}{F} and together allowing for a comprehensive reconstruction of its level and decay scheme, including widths, and predictions from theory with spectroscopic strength $C^2S$~\cite{revel20}. 
Analysis suggests that the ground state of \nuc{28}{F} is strongly dominated by $\ell=1$ contributions based on resonance width and spectroscopic strength of $C^2S=0.40(6)$.
This means that the energetically favorable configuration prefers the population of a $p$-wave resonance, exhibiting a degenerate $sd$-$pf$ shell and inversion between the $p$ and $f$ orbitals.
Comparison to a large-scale shell-model calculation utilizing the SDPF-U-MIX interaction supports the onset of the IoI in \nuc{28}{F} at $N=19$, identifying a multiplet of $J^{\pi}=(1-4)^-$ states originating from the $\pi0d_{5/2}\otimes\nu1p_{3/2}$ configuration.  

\begin{figure}[]
\centering\includegraphics[width=0.65\textwidth]{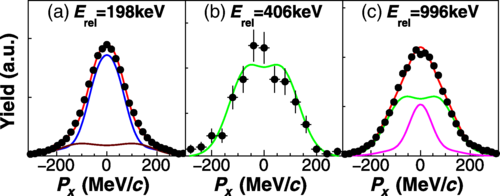}
\caption{Experimental transverse-momentum distributions of the \nuc{27}{F}$+n$ system following the \nuc{29}{F}$(-1n)$ reaction compared to eikonal-model calculations for the following~\cite{revel20}: (a) removal of a neutron with $\ell=1,3$ (respectively, blue, brown lines) for the g.s. at $E_{rel}=198$\,keV; (b) a pure $\ell=2$ distribution for the resonance at $406$\,keV (corresponding to the state at $1321$\,keV); (c) a mixture of $\ell=0,2$ (respectively, purple, green lines) for the state at $E_{rel}=996$\,keV. In (a),(c) the red line is the total fit. Reprinted figure with permission from~\cite{revel20}. Copyright 2024 by the American Physical Society.}
\label{fig:p_29f}
\end{figure}

Moreover, the neutron knockout also probes the valence-neutron ground-state configuration mixing of \nuc{29}{F}, where the momentum profile of \nuc{28}{F} states is sensitive to the angular momentum of the struck neutron.
Analysis of the measured transverse-momentum distributions ($p_x$) for the first three resonances, compared to eikonal-model reaction calculations, is shown in Fig.~\ref{fig:p_29f}~\cite{revel20}.
The ground state predominantly exhibits $\ell=1$ characteristics with a significant contribution of $79$\%, indicative of a very strong involvement of $p$ orbitals rather than the naively expected $d$-wave neutrons and their inversion with $f$ orbitals~\cite{revel20}.

A similar conclusion is indirectly drawn from a later measurement of the reaction cross-section of \nuc{29}{F} at \mbox{BigRIPS} and the Zero-Degree Spectrometer (ZDS) at \mbox{RIBF}~\cite{bagchi20}.
The measured cross section for \nuc{29}{F} is significantly larger than the $A^{2/3}$ trend, which suggests, in comparison to various shell-model and ab-initio calculations, that the \nuc{29}{F} could be interpreted as a two-neutron halo nucleus with an extended root-mean-square halo radius of $6.6$\,fm~\cite{bagchi20} and the filling of the $fp$ orbital.

\subsection{\nuc{30}{F}}
\begin{figure}[]
\centering\includegraphics[width=0.45\textwidth]{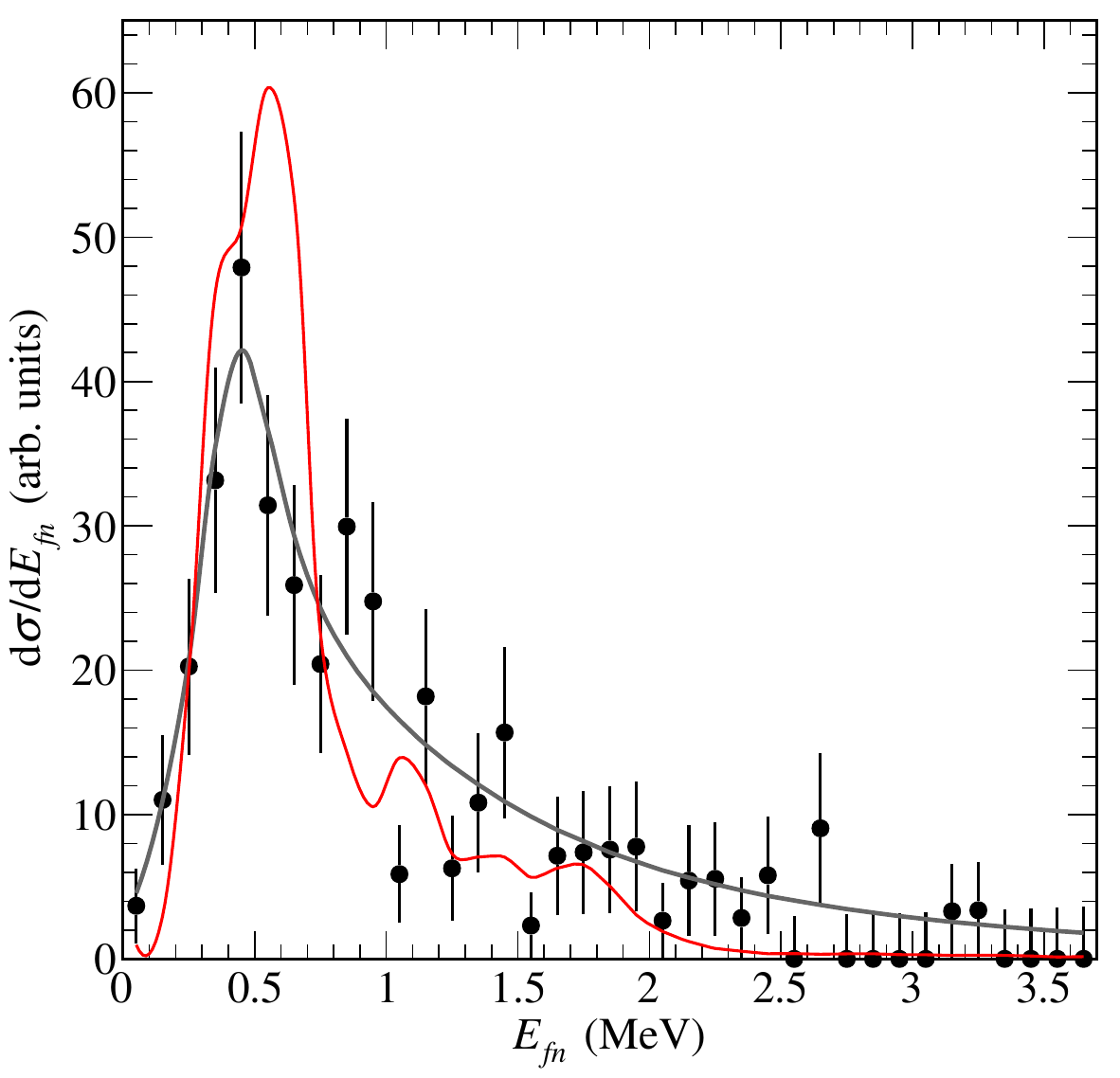}
\caption{Relative-energy spectrum of \nuc{30}{F} reconstructed
in the \nuc{31}{Ne}($p,2p$)\nuc{29}{F}$+n$ reaction. 
The data (points with $1\sigma$ stat. uncertainty) are corrected for efficiency and acceptance of the neutron detection. 
The gray curve shows the total fit, including one Breit-Wigner resonance at $E^r_{fn} = 472\pm 58 \mathrm{(stat.)} \pm 33 \mathrm{(sys.)}$\,keV together with a contribution for unresolved resonant contributions, details of the fit can be found in Ref.~\cite{kahlbow24}. 
The red curve depicts the sum of shell-model predicted \nuc{30}{F} states produced by proton knockout from \nuc{31}{Ne}, scaled by spectroscopic strength $C^2S$ and smeared by the experimental response and normalized to the data up to $2.5$\,MeV (details in Ref.~\cite{kahlbow24}). Figure adapted from Ref.~\cite{kahlbow24}.}
\label{fig:erel_30f}
\end{figure}

The observation of \nuc{30}{F}, with its $21$ neutrons, presents the first measurement surpassing $N=20$, providing pivotal direct insights into the breakdown of the $sd$-$pf$ gap.
Employing quasi-free proton-knockout on a \nuc{31}{Ne} beam populates the unbound \nuc{30}{F} which is reconstructed by means of invariant-mass spectroscopy measuring \nuc{29}{F} and the coincident decay neutron.
These data were acquired during the experimental campaign at \mbox{SAMURAI} as described above. 
Experimentally, a single resonance has been identified at $E_{rel} = 472\pm 58\mathrm{(stat.)} \pm 33\mathrm{(sys.)}$\,keV, attributed to the ground state of \nuc{30}{F}~\cite{kahlbow24}, Fig.~\ref{fig:erel_30f}.
It is noted that there are likely unresolved resonances present resulting from proton knockout from the deformed $p$-wave halo \nuc{31}{Ne}, akin to the case of \nuc{28}{F} with predicted negative-parity states (cf.~\cite{kahlbow24}).

The predicted ground-state in the \mbox{SDPF-U-MIX20} shell-model calculation is a $4^-$ state but with small overlap in the knockout reaction on \nuc{31}{Ne}.
The overlay between experimental data and the states predicted by theory, smeared by resolution and scaled by predicted strength, is shown in Fig.~\ref{fig:erel_30f}, the shapes exhibit similarity, while the theory spectrum is shifted by $-600$\,keV to adjust for slightly underpredicted binding in the theory.
The predicted $2^-$ and $3^-$ states, slightly higher in energy, have larger predicted $C^2S$ and are in better agreement with experiment, suggesting that the \nuc{30}{F} resonance associated with the ground state could be lower by approximately $200$\,keV, further details are discussed in~\cite{kahlbow24}. 

Added to the systematics of single-neutron separation energies for the F isotopes as function of neutron number with a negative value of $S_n=-472$\,keV, it is seen that the amplitude of the pairing oscillation remains essentially constant at large neutron number $N>16$, which is at variance with the empirical variation of the pairing gap following $A^{-1/2}$ and with the prediction of an increased pairing gap at low density in infinite matter~\cite{lombardo01}.
Importantly, no sharp drop in $S_n$ beyond $N=20$ is observed.
This is in stark contrast to semi-magic nuclei like \nuc{35}{P}, directly signifying that the energy gap at $N=20$ is not restored. 

\subsection{Conclusion}
The experimental campaign discussed here presents the first high-quality dataset for the most neutron-rich F isotopes.
This dataset enabled comprehensive studies of \nuc{28}{F}, \nuc{29}{F}, and, for the first time, \nuc{30}{F}, employing QFS to perform spectroscopy and investigate unbound states from various perspectives.
The results unequivocally and directly demonstrate that the neutron ``magic number'' $N=20$ is not restored in the F isotopes.
Notably, in \nuc{28}{F}, a quenched gap and inversion between the $p$ and $f$ orbital have been observed, providing clarification on prior measurements and clearly extending the IoI to $Z=9$.
The following sections will explore implications of these findings, and together with theoretical calculations discuss the degree of collectivity and indications found for a superfluid regime, as well as the formation of halo effects.

\section{The neutron-rich O isotopes}
Shy by just one proton, the neutron drip-line of the oxygen $Z=8$ isotopes is \nuc{24}{O}, while it is \nuc{31}{F} for the fluorine ($Z=9$) isotopes.
This weak binding of the O isotopes is primarily explained by the impact of repulsion contributions from the three-nucleon force, and the formation of a sub-shell closure at $N=16$~\cite{otsuka10}.
The isotopes \nuc{25-28}{O} are unbound via (multi) neutron emission.
Notably, the \nuc{26}{O}(g.s.) is unbound by only $18$\,keV~\cite{kondo16} making it almost bound and a good candidate for di-neutron emission, posing interesting nuclear-structure questions.

The \nuc{28}{O} nuclear system with eight protons and twenty neutrons is a corner stone in nuclear physics.
Doubly-magic nuclei are characterized by closed proton and neutron major shells making them a scarce but interesting species, and \nuc{28}{O} is the last potential neutron-rich doubly-magic nucleus which had not been measured before.
While most other canonical doubly-magic nuclei with nucleon numbers $2,8,20,28,50,82,126$ did not defy expectation and were found to be strong holdouts against shell evolution, the situation might be different for \nuc{28}{O} and neutron number $N=20$, resembling a structure much more similar to \nuc{29}{F} with just one unpaired additional $\pi0d_{5/2}$ proton.

The study of \nuc{28}{O} was the highlight experiment of the discussed campaign at \mbox{SAMURAI} which was possible owing to the combination of highest luminosity and neutron detection efficiency, measuring for the first time the five-body decay with four coincident fast neutrons in the reaction \nuc{29}{F}$(p,2p)$\nuc{28}{O}$\rightarrow$\nuc{24}{O}$+n+n+n+n$.
The reconstructed five-body relative energy-spectrum yields the observation of a resonance at $0.46^{+0.05}_{-0.04}$(stat.)$\pm0.02$(sys.)\,MeV which is argued to be the ground state of \nuc{28}{O}~\cite{kondo23}, while an excited state ($2^+$) has not been identified, most likely due to limited statistics.
The analysis of the four-body decay leads to the identification of the \nuc{27}{O} ground state at $1.09\pm0.04$(stat.)$\pm0.02$(sys.)\,MeV~\cite{kondo23}.
Given the low-lying ground state of \nuc{26}{O} and its high excited state at $1.28$\,MeV~\cite{kondo16} it is found that both \nuc{27,28}{O} decay dominantly in sequential decay mode through \nuc{26}{O}(g.s.).

\begin{figure}[]
\centering\includegraphics[width=0.45\textwidth]{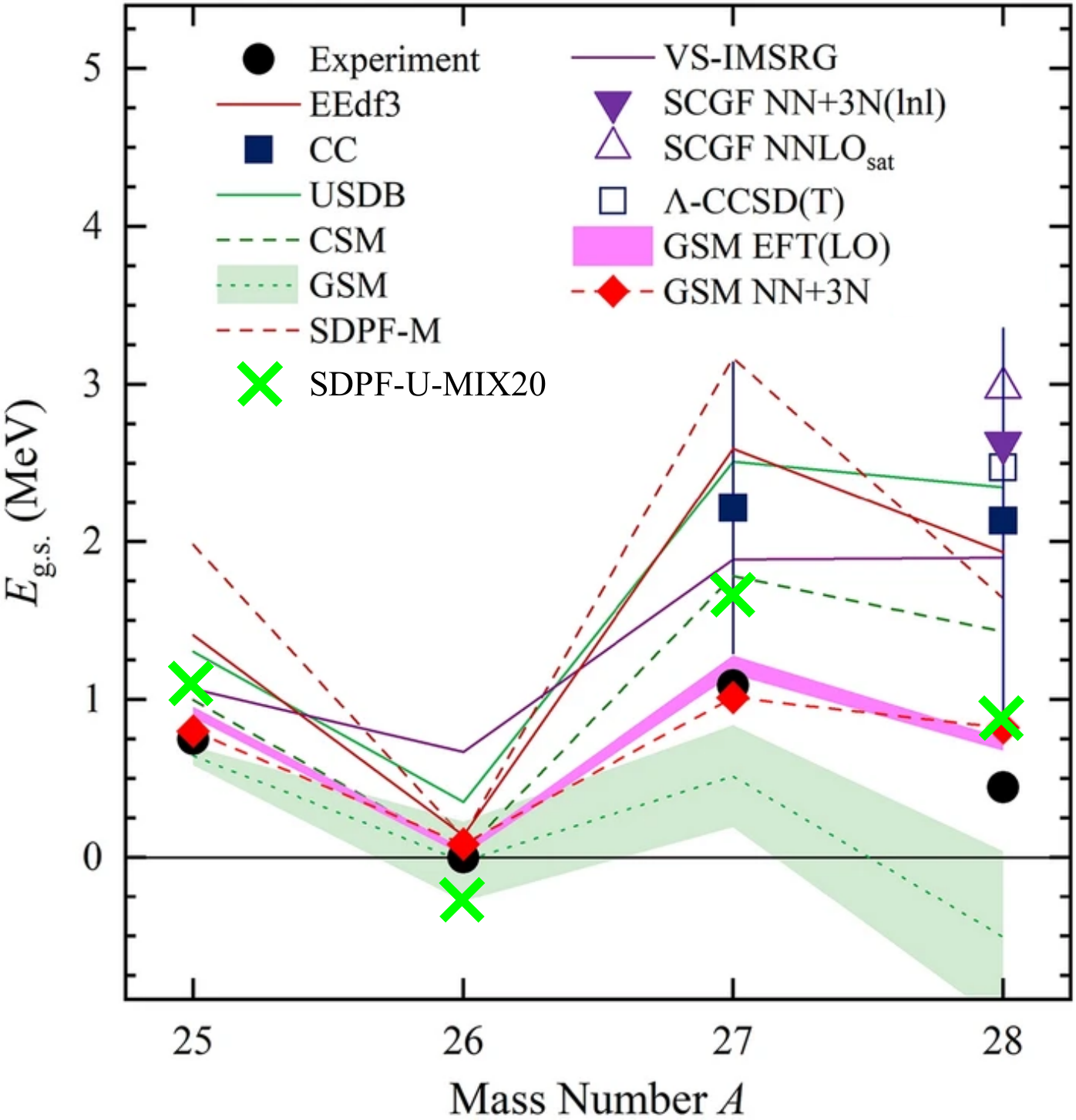}
\caption{Experimental ground-state energies of \nuc{25-28}{O} relative to \nuc{24}{O} compared to various calculations. Results are taken from \cite{kondo23}, adding the recent GSM EFT(LO) and GSM NN+3N predictions from \cite{li24}, as well as the SDPF-U-MIX20 calculation from \cite{kahlbow24}. Figure adapted from Refs.~\cite{kondo23,li24}.}
\label{fig:o_compare}
\end{figure}

While the excitation of the $2^+$ state, which is often a first signature for a large or reduced shell gap, has not been identified in \nuc{28}{O}, its nuclear structure has been analyzed by theory comparisons in that work~\cite{kondo23}:
The single-particle cross section for populating the \nuc{28}{O} resonance is obtained to be $1.36^{+0.16}_{-0.14}$(stat.)$\pm0.13$(sys.)\,mb~\cite{kondo23} and compared to distorted-wave impulse approximation calculations that determine the overlap between \nuc{29}{F} and \nuc{28}{O} in the proton knockout.
The \nuc{29}{F}(g.s.) is treated with spin-parity of $5/2^+$ as supported by the analysis of the \nuc{28}{O} momentum distribution that can be described by $d$ wave, so that mainly $d_{5/2}$ protons are knocked out from \nuc{29}{F}.
The reported $C^2S = 0.48^{+0.05}_{-0.06}\mathrm{(stat.)}\pm0.05\mathrm{(sys.)}$ is of appreciable strength, meaning that the structure, especially the neutron configuration, of \nuc{29}{F} and \nuc{28}{O} is very similar~\cite{kondo23}.
As has been established in Sec.~\ref{sec:f_isotopes} that the $N=20$ gap is quenched and orbital inversion takes place, a similar conclusion can now be drawn for \nuc{28}{O} as well.
Additionally, the experimental ground-state energies for the O isotopic chain are compared to predictions from various shell-model and ab-initio calculations in Fig.~\ref{fig:o_compare} (details in Sec.~\ref{sec:theory}), supporting the strong indications that \nuc{28}{O} is not a doubly-magic nucleus.

\section{Nuclear structure theory}
\label{sec:theory}
Witnessed by the experimental results above, nuclear structure undergoes significant evolution in regions characterized by extreme neutron-proton asymmetry.
This evolution plays a pivotal role in determining various aspects such as drip lines, single-particle structure, deformation, collectivity, and numerous other properties of the nuclear many-body system.
Consequently, the study of neutron-rich nuclei becomes particularly crucial for understanding the underlying nuclear interaction.
Significant efforts in nuclear theory are directed to establish a description of nuclear interactions that can universally describe a large region in the chart of nuclides.
Ab-initio approaches are especially focused on achieving a microscopic derivation from first principles and QCD symmetries.

\subsection{Shell-Model calculations}
Beyond the pure single-particle interpretation, the nuclear structure and energy levels are significantly influenced by ``residual'' interactions, often denoted as effective single-particle energies, comprising two-body and higher multi-nucleon contributions.
In shell-model approaches, trends of residual two-body matrix elements are determined through fits to experimental data.

The interaction used to compare to the discussed F and O experimental results is the so-called \mbox{SDPF-U-MIX} interaction, which specifically encompasses the $sd$-$pf$ neutron model space~\cite{caurier14}.
It effectively describes a wide range of nuclei spanning from $sd$ to $pf$ regions around \nuc{32}{Mg} up to the Ca isotopes, especially pivotal for understanding nuclei around $N=20$ in the IoI.

\begin{figure}[]
\centering\includegraphics[width=0.45\textwidth]{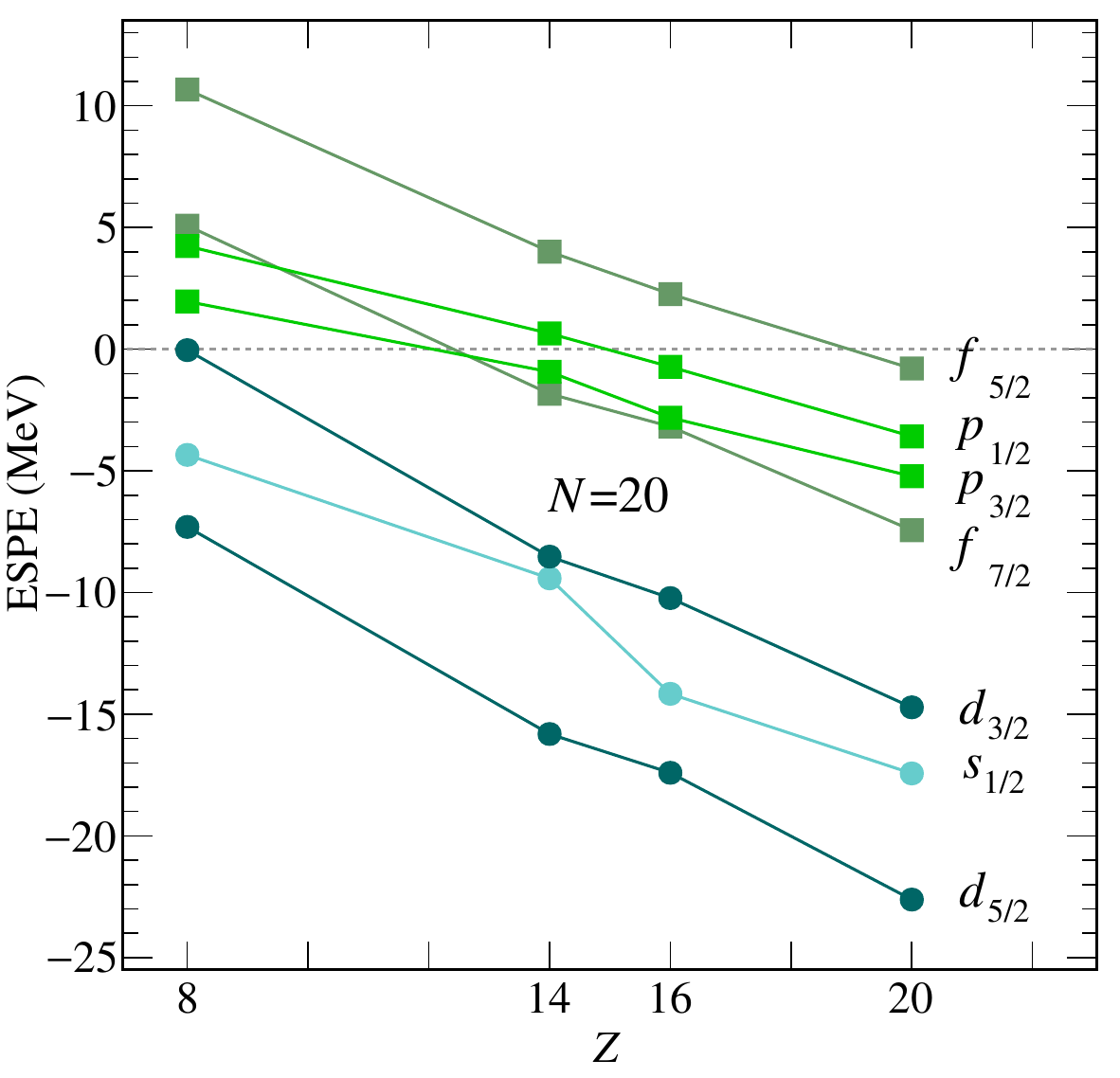}
\caption{Evolution of the neutron Effective Single-Particle Energies (ESPE) of the $sd$ and $fp$ orbitals as a function of the proton number $Z$ for $N=20$ isotones, calculated with the SDPF-U-MIX20 interaction. The $N=20$ shell gap, normally 
formed between the $0d_{3/2}$ and $0f_{7/2}$ orbitals, is present from $Z=20$ down to $Z=14$ but collapses completely at $Z=8$. There, we also observe an inversion of orbitals, with the $1p_{3/2}$ and $1p_{1/2}$ ones below $0f_{7/2}$. Figure taken from Ref.~\cite{kahlbow24}.}
\label{fig:theo_n20} 
\end{figure}

The neutron-proton monopole component, mainly of central and tensor form, leads to the increase of the $N=20$ gap with increasing proton number as indicated in Fig.~\ref{fig:theo_n20}, using the \mbox{SDPF-U-MIX20} interaction~\cite{kahlbow24}, a behavior observed in many regions of the chart of nuclides caused by the monopole term.
Importantly, the interplay with (multipole) correlations, primarily of pairing and quadrupole character, dictates the nuclear-structure physics.
In the case of the neutron-rich F and O isotopes within the weak-binding region, it becomes energetically favorable for the valence neutrons to occupy orbitals beyond the traditionally assigned ones, such as the $p$ and $f$ orbitals above the (quenched) $N=20$ gap.

As shown in Fig.~\ref{fig:theo_n20}, the $N=20$ gap diminishes from approximately $15$\,MeV at $Z=14$ to about $2$\,MeV at $Z=8$, where the $sd$ and $pf$ shells become essentially degenerate, and the $p$ orbitals become more bound than the $f$ orbital.
The results of this large-scale shell-model calculation are in good agreement with experimental data for the F isotopes, and reproducing the trend of the single-neutron separation energy $S_n$ up to the new result at $N=21$~\cite{kahlbow24}, as well as the spectroscopy of \nuc{28}{F} and \nuc{30}{F}~\cite{revel20,kahlbow24}.
The neutron occupancy of the $0f_{7/2}$ and $1p_{3/2}$ orbitals is predicted to be $0.2$ and $0.8$, respectively, in \nuc{29}{F}, and $0.8$ and $2.1$, respectively, in \nuc{31}{F}.
Turning to the O isotopes, the $S_n$ and ground-state energies are also fairly well reproduced, as can be seen in Fig.~\ref{fig:o_compare}, although all ground states are unbound and contributions from continuum degrees of freedom should be relevant.

Components from multi-nucleon forces and continuum contributions are traditionally not fully and explicitly incorporated in the shell-model interactions.
However, it is known that the three-nucleon force is crucial for reproducing the O drip line~\cite{otsuka10}. 
Similarly, it appears natural that continuum coupling affects the interaction of unbound nuclei and those close the drip line, as discussed in Sec.~\ref{sec:continuum}.
These contributions are only indirectly and partially accounted for in the shell-model approach through fitting nuclear data of other neutron-rich nuclei and their impact on two-body matrix elements.
Nevertheless, the shell-model results exhibit good agreement with data, potentially even outperforming the ab-initio results, cf. Fig.~\ref{fig:o_compare}.

\subsection{Ab-initio calculations}
First-principle ab-initio approaches seek to derive the effective nuclear interaction from fundamental QCD symmetries that in the highest possible resolution describe the effective nuclear interaction.
One very successful method among these ab-initio theories are chiral Effective Field Theories (EFT), which are constrained by a limited number of coupling constants to systematically expand the theory and quantify uncertainties. 
The EFT interactions serve as input to the Hamiltonian which is then solved using many-body methods, only recently extending to the drip line of medium-mass nuclei including O and F~\cite{hergert20,stroberg21}.

The recent results for the oxygen isotopic chain are compared to variety of ab-initio results utilizing different interactions and many-body methods, as shown in Fig.~\ref{fig:o_compare} with calculations compiled in~\cite{kondo23}.
It can be seen that essentially all calculations over-predict the ground-state energies of \nuc{27}{O} and \nuc{28}{O} by approximately $1$ to $2$\,MeV.
The Gamow Shell-Model calculation (GSM) behaves differently, it instead exhibits overbinding (green band), and later added GSM calculations (GSM EFT and NN$+3$N) show very good agreement~\cite{li24}.
The authors of Ref.~\cite{kondo23} discuss an example using the \mbox{EEdf3} interaction in a large-scale shell-model calculation demonstrating the breakdown of the $N=20$ magic number, despite the energy offset of about $1$\,MeV.
While all calculations include terms for three-nucleon forces (except the GSM), no particular interaction appears to be favored.
Additionally, the application of different many-body methods contributes to variations in results; for instance, the Coupled Cluster calculation ($\Lambda$-CCSD(T)) utilizing the same NNLO$_{sat}$ interaction as the Self-consistent Green's Function (SCGF) calculation yields results differing by approximately $0.5$\,MeV for \nuc{28}{O}. 
Furthermore, the result obtained using the same SCGF method but with a different interaction, namely NN$+3$N(lnl), falls between these values.

Ab-initio calculations of the neutron-rich F isotopes remain challenging, especially incorporating continuum coupling.
A previous valence-space in-medium similarity renormalization group (VS-IMSRG) calculation however shows good agreement with recent experimental data in terms of $S_n$ values (up to $A=29$), while absolute magnitudes of g.\,s. energies appear to be shifted~\cite{stroberg16,hu20}.
More is discussed in context of continuum coupling in the next section.

Ab-initio calculations allow to gain a fundamental and systematic understanding of the nuclear interaction and many-body system.
In Ref.~\cite{kondo23}, the authors argue that \nuc{27,28}{O} are valuable anchors for theoretical approaches based on $\chi$EFT.
In particular, the authors explore efficient and fast many-body methods using emulator techniques with eigenvector continuation and history matching, as well as uncertainty quantification using Bayesian theory based on the Coupled Cluster calculation.

\subsection{Continuum coupling}
\label{sec:continuum}
\begin{figure}[]
\centering\includegraphics[width=0.7\textwidth]{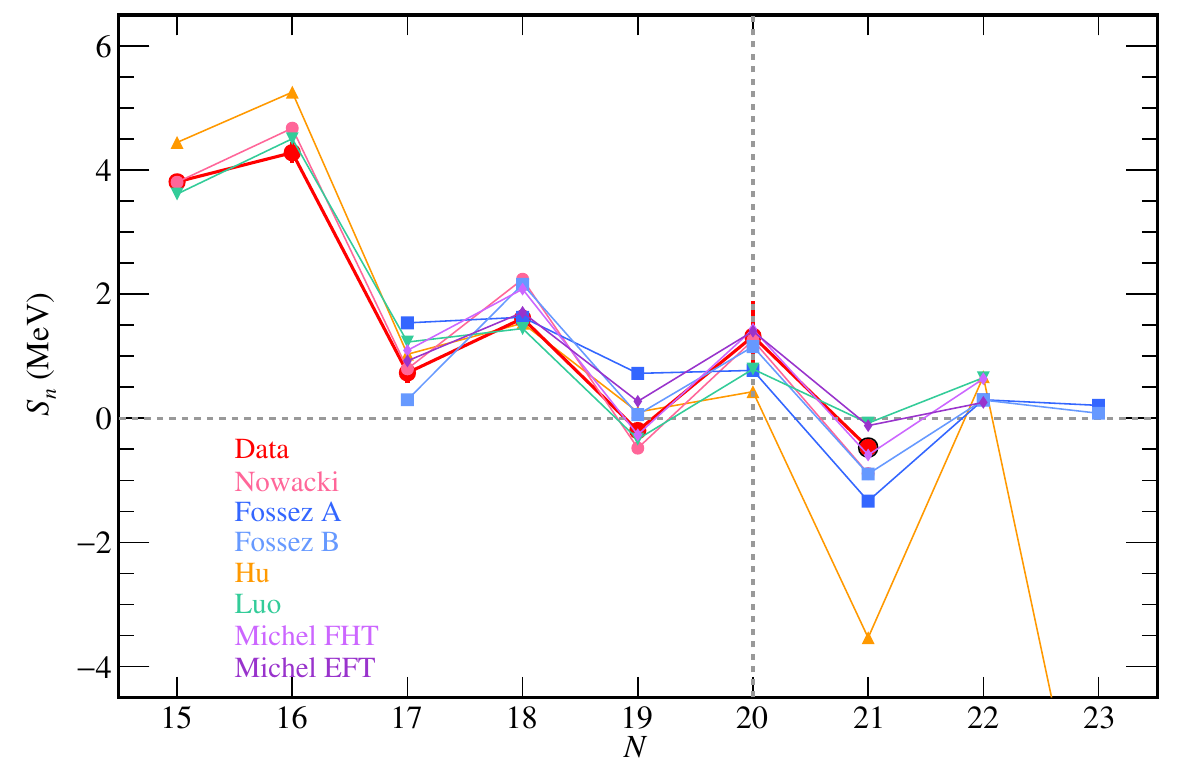}
\caption{$S_n$ values for F isotopes as function of neutron number $N$~\cite{kahlbow24}, comparing data (red) to results from \mbox{SDPF-U-MIX20} shell-model calculation (``Nowacki'')~\cite{kahlbow24} and various calculations that explicitly include continuum degrees of freedom: density renormalization group method (azure, ``Fossez A \& B'')~\cite{fossez22}, ab-initio Gamow shell model (orange, ``Hu'')~\cite{hu20}, shell-model embedded in the continuum (green, ``Luo'')~\cite{luo02}, and Gamow shell model (pink, ``Michel FHT \& EFT'')~\cite{michel20}. Figure taken from Ref.~\cite{kahlbow24}.}
\label{fig:sn_continuum} 
\end{figure}

As nuclei approach the drip line, where they become very weakly bound or unbound, continuum degrees of freedom should play a crucial role in coupling to the system.
The studied O and F isotopes might exhibit pronounced competing effects of deformation and continuum coupling.
Theoretical challenges arise in accounting for continuum coupling due to the large amount of possible scattering states.
Established methods include the Gamow Shell Model, the Density Renormalization Group (DMRG) method, or the Shell Model embedded in the continuum method.
Recent efforts have focused on incorporating continuum effects into structure calculations of the O and F isotopes, including within ab-initio approaches.
Figure~\ref{fig:sn_continuum} shows a compilation of results for $S_n$ in the F chain by various theories that incorporate continuum degrees of freedom explicitly~\cite{kahlbow24} (and Refs. therein).

Efforts to extend the phenomenological shell model approach by incorporating continuum correlations are explored in the shell-model embedded in the continuum (\mbox{SMEC}) calculations.
The work Ref.~\cite{luo02} identifies a description of additional terms in the shell model calculations that lead to significant attenuation of binding energies.
This reduction depends strongly on the position of the states relative to the neutron emission threshold energy and the degree of asymmetry.
However, the magnitude of the continuum coupling strength remained a parameter in this initial investigation and was tuned.

The work of Hu \textit{et al.}~\cite{hu20} presents the first ab-initio construction of an $sdpf$ multi-shell Hamiltonian within the Gamow Shell Model (GSM), based on the chiral NNLO$_{\mathrm{opt}}$ interaction.
A key challenge in such calculations lies in the choice of the basis states, particularly when considering continuum for which not only bound, but resonant and scattering states need to be included.
The authors employ the Gamow Hartree-Fock approach to generate the Berggren basis, which treats bound, resonance, and scattering states on an equal footing in the complex momentum plane~\cite{hu20}.
Incorporating many-body correlations, $pf$-shell configurations, and continuum degrees of freedom, the authors present results for ground-state and excited-state energies for neutron-rich O and F isotopes.
In case of the unbound O isotopes, inclusion of the continuum tends to bind the nuclei stronger by about $1$\,MeV compared to the same calculation without continuum, improving agreement with experimental results.
However, for F isotopes with $A=30,32$, including the continuum appears to decrease binding strongly, posing a challenge in this continuum calculation~\cite{hu20}.
The GSM calculation by Michel et al.~\cite{michel20} examines two interactions, a phenomenological one and one derived from effective field theory.
While the EFT one over binds, the results from the other calculation seem to match experimental results well, but it is noted that this is a locally tuned interaction and limited to a \nuc{24}{O} core, two $d_{5/2}$ and $s_{1/2}$ proton orbitals, and $d_{3/2}$, $f_{7/2}$ and $p_{3/2}$ neutron orbitals.

Another recent work, by Fossez and Rotureau~\cite{fossez22}, applies the DMRG method to study the spectroscopic properties of the neutron-rich F isotopes $A=25-33$, i.\,e., open quantum systems in the $sd$-$pf$ model space.
This work uses a single-particle potential with a simplified interaction based on halo effective field theory and the Berggren basis.
The authors identify two competing effects -- weak binding and deformation (discussed below) -- with deformation primarily driven by the coupling of the neutron $1p_{3/2}$ and $0f_{7/2}$ orbitals.
They argue that the $1p_{3/2}$ orbital is favored over the $0d_{3/2}$ due to continuum coupling, highlighting the significance of the position of the $p_{3/2}$ in neutron-rich F isotopes.
In terms of $S_n$ values, the results tend to over bind the unbound \nuc{28,32}{F} nuclei.
For the detailed spectroscopy of configuration mixing and excited states, two versions of the interaction are tested, one better applicable to \nuc{25-29}{F} and the other to \nuc{30-33}{F}.
While also finding the IoI, the results show a slight preference for the first $1/2^+$ state to be lower in energy compared to the $5/2^+$ state for \nuc{29-33}{F}~\cite{fossez22}, which seems to be at odds with experiment in the case of \nuc{29}{F}~\cite{kondo23}.
Additionally, this works predicts a deformed $2n$ halo ground state in \nuc{29}{F}, with two neutrons likely occupying the $p_{3/2}$ orbital.
A similar effect is predicted for \nuc{31}{F}, characterized by a strong $\ell=1$ contribution, resulting in the loss of single-particle character.
The emergence of deformation in this region of the chart of nuclides is also discussed in the next section.

Continuum degrees of freedom remain a challenge for theory to fully absorb.
The results presented in Fig.~\ref{fig:sn_continuum} are often only locally defined with restricted interactions that would often not be able to predict properties of surrounding isotopes, in contrast to large-scale shell-model calculations without continuum.
Given the comparison of data and shell model for F, it seems that continuum contributes only moderately.

\subsection{Deformation}
Magic nuclei typically exhibit a spherical shape, contrasting with most other nuclei that do not show this symmetry.
As discussed above, the IoI is characterized by a loss of ``magicity'', indicating the emergence of deformation in $N=20$ nuclei as a result of shell evolution. 
In the shell-model picture, deformation is attributed to the strength of multipole interactions, particularly the interplay between quadrupole and pairing interactions, which drive shell evolution.
The work~\cite{fossez22} predicts deformation to develop in the neutron-rich F isotopes in competition with impact of continuum coupling.
Current efforts are direct to understanding deformation from first principles.

\begin{figure}[]
\centering\includegraphics[width=0.7\textwidth]{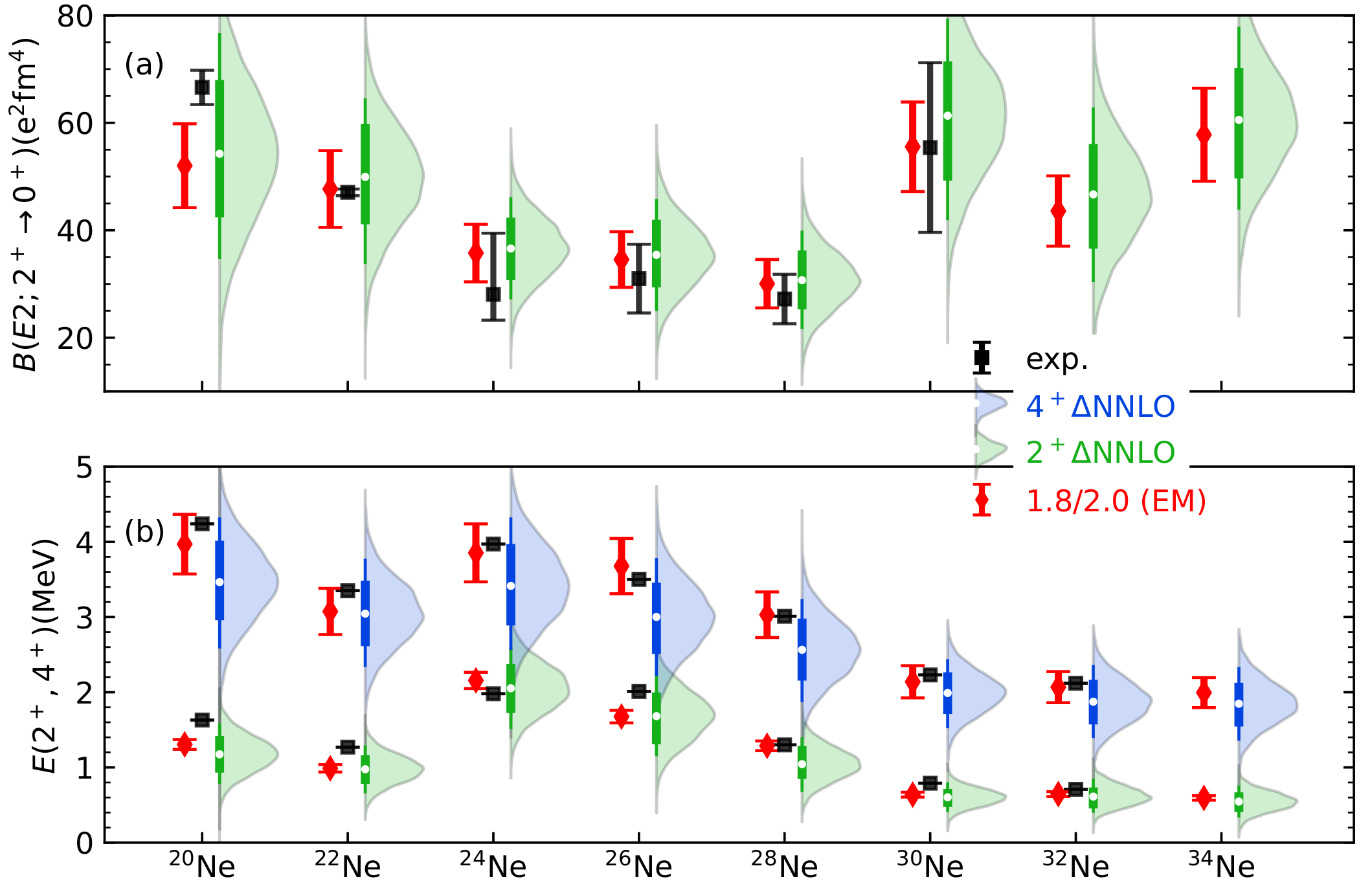}
\caption{(a) Electric quadrupole transition strengths from the first excited $2^+$ state to the ground-state, (b) energies of the lowest $2^+$ and $4^+$ states in the even nuclei \nuc{20-34}{Ne} using EFT. Theoretical results are computed using angular-momentum projected coupled-cluster. Point predictions using the interaction $1.8/2.0$(EM) are shown together with full posterior predictive distributions using the delta-full NNLO interaction ensemble ($\Delta$NNLO). Figure taken from Ref.~\cite{sun24} Copyright CC BY 4.0.}
\label{fig:abinitio_deformation} 
\end{figure}

Throughout the IoI, nuclei are found to be deformed and exhibit shape coexistence, with a notable example being \nuc{32}{Mg}.
Explaining deformation and calculating rotational-band structures, as well as collective electromagnetic transition strength poses a challenge for ab-initio calculations, a topic that has only recently been addressed~\cite{tsunoda20,miyagi20,sun24,ekstroem23}.
In the recent study in Ref.~\cite{sun24}, the structure of even neutron-rich Ne and Mg isotopes was investigated, focusing on the strength of different factors causing deformation, by developing a unified and non-perturbative quantum many-body framework that captures both short- and long-range correlations starting from ab-initio theory, following works of~\cite{novario20,hagen22} calculating charge radii and structure in Ne and Mg isotopes.
Utilizing a $\chi$EFT interaction ($1.8/2.0$ EM), incorporating three-nucleon forces, and accounting for uncertainties through Bayesian analysis, most neutron-rich Ne and Mg isotopes are predicted to exhibit axial deformation, behaving as rigid rotors.
This conclusion is based on the ratio of the first $4^+/2^+$ states, yielding values around $10/3$ for isotopes \nuc{20,30}{Ne} and \nuc{34}{Mg} as shown in Fig.~\ref{fig:abinitio_deformation}~\cite{sun24}.
The results for low-lying collective states and the electromagnetic quadrupole transitions agree well with experimental results within uncertainty.
In \nuc{30}{Ne} coexisting spherical and deformed shapes are found, which is another indication for the breakdown of the magicity at $N=20$ close to \nuc{28}{O}~\cite{sun24}.
Similarly, strong deformation and collectivity is predicted for \nuc{32}{Ne} and \nuc{34}{Ne}, while isotopes \nuc{34-38}{Mg} are expected to have prolate deformed ground states, consistent with experimental data.
The most neutron-rich Mg isotope studied so far, \nuc{40}{Mg}~\cite{crawford19,tsunoda20}, indicates shape coexistence based on comparison to theory, with a prolate ground-state rotational band and an excited oblate band~\cite{sun24}.
The results demonstrate that the inclusion of short- and long-range correlations on top of an axially deformed reference state enables to accurately capture quadrupole collectivity and that deformation and shape-coexistence emerge in a multiscale setting capturing both small excitation energies and large total binding energies.

This ab-initio calculation represents the first attempt to establish a link between $\chi$EFT and deformation~\cite{ekstroem23,sun24}, employing sensitivity studies and additionally using emulator techniques to help expedite calculations.
It is concluded that, from a microscopic perspective, the singlet $S$-wave short-range contact term serves as the primary driving mechanism of deformation as related to pairing, contributing approximately $50$\% for \nuc{20,32}{Ne} and \nuc{34}{Mg}.
The framework developed in Ref.~\cite{sun24} successfully allows to model and understand multi-scale phenomena from a microscopic perspective.

\subsection{Conclusion}
State-of-the-art calculations of the neutron-rich F and O isotopes around $N=20$ predict significant shell evolution, a departure from the classical shell model, driven by both deformation and continuum coupling.
This evolution in nuclear structure has been substantiated experimentally, extending the understanding of the IoI to include the F and O isotopes.
The presented large-scale shell-model calculation utilizing the SDPF-U-MIX interaction demonstrates better agreement with experimental ground-state energies along isotopic chains within the IoI compared to various ab-initio calculations employing different microscopic interactions.
However, the neutron-rich F and O isotopes, in particular, present a rigorous test for modern nuclear structure theories.
They offer insights into crucial questions concerning three-nucleon forces, the influence of continuum coupling, superfluidity versus magicity, and prompting discussions on deformation from a microscopic perspective.

\section{Impact}
The cutting-edge experiments at \mbox{SAMURAI} at RIBF, particularly utilizing the \mbox{MINOS} LH$_2$ target device, have yielded invaluable insight into the nuclear structure of open quantum systems, particularly focusing on the most neutron-rich O and F isotopes.
These experiments serve as a corner stone for advancing nuclear theory, with large-scale shell-model calculations adeptly describing the ``Island of Inversion''.
Ab-initio calculations are rigorously tested against these new results, offering an opportunity to develop the most advanced microscopic descriptions of continuum coupling and emergent symmetries.

\paragraph{Superfluidity}
The shell-model calculation discussed in Sec.~\ref{sec:theory} indicates a close proximity of all neutron orbits, with the proton $Z=8$ core likely preserved, resulting in interesting properties for the collectivity (quadrupole, pairing) of these nuclei~\cite{kahlbow24}. 
According to these calculations, \nuc{28}{O} exhibits 97\% of neutron pairs coupled to $J=0$ (seniority $0$), with 50\% in a closed-shell configuration and 47\% of pairs involved in $sd$-to-$fp$ excitations, mostly between the $d_{3/2}$ and the $p_{3/2}$ orbitals. 
Similarly, \nuc{29}{F} shows approximately 70\% of $J=0$ pairs, with more than 60\% of these pairs scattering from the $sd$ to $pf$ orbits~\cite{kahlbow24}.
Such a regime in which pairs of nucleons equivalently occupy nearby orbitals is usually defined as superfluidity, which Ref.~\cite{kahlbow24} proclaims in this region of the chart of nuclei for the first time.
A classic example of superfluid behavior is the tin isotopic chain between $N=50-82$.
Additionally, this behavior is characterized by almost constant pairing oscillations, as observed in the F isotopes. 
This suggests that, instead of classical nuclear shell structure and magicity, these nuclei exhibit a different phase, namely superfluidity, which could be a signature of neutron-saturated systems.
The looming question pertains to the nature of the superfluid regime, whether it aligns with the Cooper pair (BCS) or Bose Einstein Condensate (BEC) type, where BEC is characterized by close and compact neutron pair sizes as opposed to the BCS type with large-correlation length neutron pairs.
While potentially preferring the BEC type with small-size neutron pairs because of mixing between weakly bound orbits of different parities, future experiments and theoretical studies are urged to identify and extract signatures for neutron correlations using, for example, direct studies of neutron decay correlations.

\paragraph{Halo}
The valence neutron configuration in \nuc{29}{F}, as experimentally confirmed by the study of \nuc{28}{F}, favors the occupancy of the $p$ orbital.
Given the presence of these low-$\ell$ orbitals, their high occupancy, and their relatively weak binding, it is likely that two-neutron halo structures could develop in \nuc{29}{F}, as well as in \nuc{31}{F}~\cite{kahlbow24}. 
The shell-model calculation presented in Ref.~\cite{kahlbow24} predicts a significant occupancy of the $p_{3/2}$ orbital by $0.8$ neutrons, along with a low $S_{2n}$ value of $776$\,keV for \nuc{29}{F}, compared to the experimental value of $1130(540)$\,keV, and $2.1$ neutrons and $S_{2n}=32$\,keV for \nuc{31}{F}.
The work by Fortunato \textit{et al.}~\cite{fortunato20} predicts a \nuc{29}{F} ground state valence structure with 57.5\% $(1p_{3/2})^2$, 28.1\% $(0d_{3/2})^2$, and 6\% of $(0f_{7/2})^2$, adjusting to reproduce~\cite{revel20}, resulting in a non-negligible increase in the matter radius with respect to the \nuc{27}{F} core of $\Delta R=0.192$\,fm, which may indicate moderate halo formation in \nuc{29}{F}. 
This appears to predict smaller $p_{3/2}$ occupancy and smaller radius than the work~\cite{bagchi20}.  
The authors further state that intruder components tend to favor a dineutron configuration due to the mixing.
A two-neutron halo structure in \nuc{29}{F} is also predicted by~\cite{fossez22}, albeit with a $1/2^+$ ground state and slight deformation.
Conversely, proton knockout on \nuc{29}{F} indicates the removal of $d$-wave protons, suggesting a $5/2^+$ ground state~\cite{kondo23}. 
Interestingly, that work predicts additional one-neutron halo states in the excited-state spectrum, with the exception of the $5/2^+$ state.
Similarly, \nuc{31}{F} is predicted to exhibit several deformed two-neutron halo configurations in the ground state and the three lowest lying states~\cite{fossez22}.
The Gamow Shell Model calculation by Michel et al.~\cite{michel20} also predicts a strong two-neutron halo in \nuc{31}{F} with a significantly enhanced RMS radius even compared to \nuc{29}{F}, with two neutrons orbiting above a \nuc{29}{F} involving $p$, $d$, $f$ neutron states.
Experimental indications of \nuc{29}{F} showing signs of a halo nucleus, namely an extended matter radius in addition to weak binding, come from a measurement of the interaction cross section of \nuc{29}{F} on a carbon target using the \mbox{BigRIPS} fragment separator and Zero-Degree Spectrometer at RIBF~\cite{bagchi20}.  
To elucidate the halo structure of the \nuc{29}{F} ground state, Coulomb excitation experiments to obtain the electromagnetic response by $E1$ strength are ideal tools~\cite{fortunato20}.
Such experiments are planned for the near future.

\paragraph{Outlook}
The high-quality data presented open the possibility to study a wealth of nuclear structure physics in the neutron-rich F isotopes.
Firstly, further experiments are warranted to extend detailed studies beyond \nuc{29}{F} and \nuc{30}{F}.
Secondly, detailed studies of \nuc{29}{F} are necessary to determine its spin-parity, neutron correlations, and to directly study halo structure and deformation.
Following the observation of \nuc{28}{O}, its structure remains largely unclear, necessitating additional measurements to attempt to detect the $2^+$ state, neutron correlations or, in combination with \nuc{29}{F}, the overlap in knockout reactions.
Some of these experiments can be performed using existing setups, while others may require higher luminosities that will become available in the coming years at facilities such as FRIB, RIBF, or GSI-FAIR.

\section*{Acknowledgment}
I thank G. Hagen and Y. Kondo for their contributions to the workshop and valuable discussions in preparing this review article.

\bibliographystyle{unsrtnat}
\bibliography{IoI_ptep}

\end{document}